\documentclass{emulateapj}
\usepackage{times}

\newcommand{\source}{\hbox{Cen\,A}}
\newcommand{\ntof}{\hbox{NGC\,315}}
\newcommand{\chandra}{\textit{Chandra}}

\begin{document}

\slugcomment{Dec 13th 2007: Accepted for publication in ApJ Letters}

\shorttitle{Cen A X-ray knottiness}
\shortauthors{D.M.~Worrall et al.}

\title{Where Centaurus A gets its X-ray knottiness}

\author{
D.M.~Worrall\altaffilmark{1,2}, 
M.~Birkinshaw\altaffilmark{1,2},
R.P.~Kraft\altaffilmark{2},
G.R.~Sivakoff\altaffilmark{3},
A.~Jord\'{a}n\altaffilmark{2},
M.J.~Hardcastle\altaffilmark{4},
N.J.~Brassington\altaffilmark{2},
J.H.~Croston\altaffilmark{4},
D.A.~Evans\altaffilmark{2},
W.R.~Forman\altaffilmark{2},
W.E.~Harris\altaffilmark{5},
C.~Jones\altaffilmark{2},
A.M.~Juett\altaffilmark{6}, 
S.S.~Murray\altaffilmark{2},
P.E.J.~Nulsen\altaffilmark{2},
S.~Raychaudhury\altaffilmark{7,2},
C.L.~Sarazin\altaffilmark{6},
K.A.~Woodley\altaffilmark{5}
}
\altaffiltext{1}{Department of Physics, 
University of Bristol, 
Tyndall Avenue, 
Bristol\ \ BS8 1TL, UK}
\altaffiltext{2}{Harvard-Smithsonian Center for Astrophysics, 
60 Garden Street, 
Cambridge, MA\ \ 02138}
\altaffiltext{3}{
Department of Astronomy,
The Ohio State University,
4055 McPherson Laboratory,
140 W.~18th Avenue, 
Columbus, OH\ \ 43210}
\altaffiltext{4}{School of Physics, Astronomy \& Mathematics,
University of Hertfordshire, College Lane, Hatfield\ \ AL10 9AB, UK}
\altaffiltext{5}{
Department of Physics and Astronomy,
McMaster University,
Hamilton, ON\ \ L8S 4M1, Canada}
\altaffiltext{6}{
Department of Astronomy,
University of Virginia,
P. O. Box 400325,
Charlottesville, VA\ \ 22904
}
\altaffiltext{7}{
School of Physics and Astronomy,
University of Birmingham,
Edgbaston, Birmingham\ \ B15 2TT, UK
}


\begin{abstract}

We report an X-ray spectral study of the transverse structure of the
Centaurus~A jet using new data from the \chandra\ \source\ Very Large
Project.  We find that the spectrum steepens with increasing distance
from the jet axis, and that this steepening can be attributed to a
change in the average spectrum of the knotty emission.  Such a trend
is unexpected if the knots are predominantly a surface feature
residing in a shear layer between faster and slower flows.  We suggest
that the spectral steepening of the knot emission as a function of
distance from the jet axis is due to knot migration, implying a
component of transverse motion of knots within the flow.

\end{abstract}


\section{Introduction}
\label{sec:intro}

In relatively few extragalactic radio jets can the kinematics of the
flow on kpc scales be studied with any degree of certainty.  Most
arguments are indirect, relying either on beaming statistics for
samples \citep[e.g.,][]{hard03a} or, for powerful jets oriented along
the line of sight, on models relying on an assumed inverse-Compton
origin for the X-ray emission \citep[e.g.,][]{schwartz06}.  For the
few low-power radio galaxies with heavily studied, straight, radio
jets and counterjets, kinematic models have been constructed based on
the jet-counterjet asymmetry \citep[e.g.,][]{canvin05}.  These models
have supported jet deceleration through mass entrainment, and ongoing
work has already shown consistency between the density and pressure
model inferred from the jet and the properties of the external
X-ray-emitting gas for 3C\,31 \citep{laing02}.

In models of mass entrainment, the outer parts (``sheath'') are
decelerated before the inner (``spine''), consistent with a range of
observational evidence at radio frequencies \citep{laing96}.  Applied
to more central regions, the consequence that emission from a slower
sheath becomes relatively more important in jets at larger angle to
the line of sight then resolves difficulties in models that unify BL
Lac objects with low power radio galaxies
\citep[e.g.,][]{chiaberge00}.

\chandra\ has not only allowed confirmation that X-ray synchrotron
emission in low-power jets is common \citep{worrall01}, but has also
resolved transverse structure in the nearest and brightest sources.  A
particularly interesting example is \ntof\ \citep{worrall07}.  Here
the diffuse emission contains a knotty structure in the radio and
X-ray that appears to describe an oscillatory filament.  Although the
structure could be the result of a chance superposition of
non-axisymmetric knots, the level of coherence led us to suggest that
the knots might be predominantly a surface feature residing in the
shear layer between the spine and sheath.  If this latter
interpretation is correct, we might expect the X-ray spectra of the
knots to be similar across the transverse width of the jet.

In the case of \ntof\ the distinct knotty emission is only about 10\%
of the total emission in X-rays and radio along the $\sim 2.5$ kpc of
projected jet length over which it is detected, and with a source
distance of $\sim 70$~Mpc the observations did not allow us to measure
separate X-ray spectra for the knots and diffuse emission.  At
3.7~Mpc\footnote{The average of five distance indicators, see \S6 in
\citet{ferrarese07}} (1~arcmin is $\sim 1.1$~kpc projected length),
Centaurus~A is a much closer example of a low-power radio galaxy whose
inner jet shows bright resolved X-ray knots and diffuse emission over
a similar projected length scale to \ntof\ \citep[e.g.][]{kraft02}.
Here we report an X-ray spectral study of the transverse structure of
the jet, and in particular we investigate whether the \source\ knots
are likely to occupy a shear layer between faster inner and slower
outer flows.

\section{Observations \& analysis}
\label{sec:obs}

\begin{figure*}
\centering
\includegraphics[width=1.9\columnwidth,clip=true]{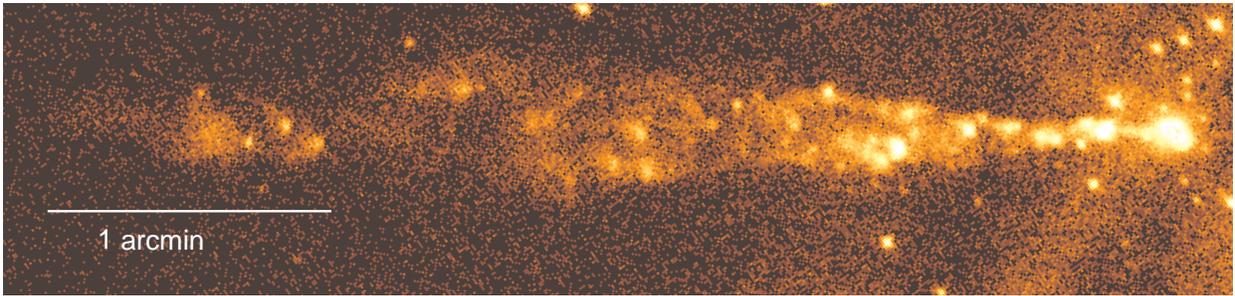}
\caption{0.8-3-keV unsmoothed rotated image of the six
\chandra\ exposures comprising the \source-VLP,
with $0.492 \times 0.492$ arcsec$^2$  pixels. }
\label{fig:jet}
\end{figure*}

\begin{figure}
\centering
\includegraphics[width=1.0\columnwidth,clip=true]{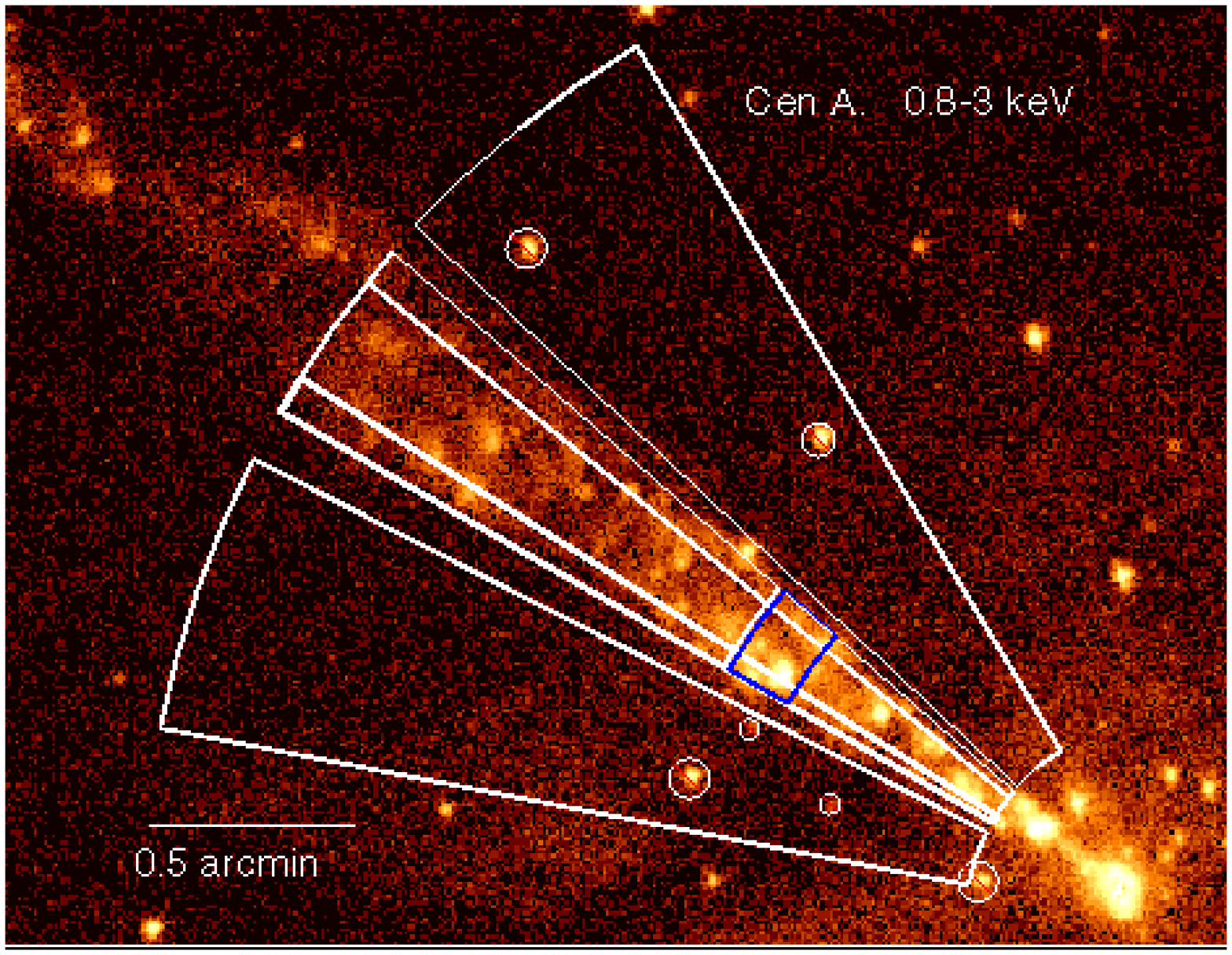}
\caption{Image from Fig.~\ref{fig:jet} with regions superposed, and
rotated to the sky orientation. The regions, called the spine, sheath and
background, working
outwards from the jet axis, are pie slices from a base position
of R.A.~$=13^{\rm h} 25^{\rm m} 26\fs98$, Dec.~$=-43\arcdeg01\arcmin14\farcs06$
(not the core) running from position angle (pa) 51\fdg2--57\fdg8 for the spine,
pa 49$^\circ$--51\fdg2 and 57\fdg8--60$^\circ$ for the sheath, and 
pa 31$^\circ$--47$^\circ$ and 63$^\circ$--79$^\circ$ for the
background with the exclusion regions shown.  From the core the inner
spine and sheath regions are at a distance of $21''$ to $66''$, and the
outer region runs from $66''$ to $141''$.  We call the region marked in blue,
near the outer extent of the inner region, the ``bright''
region.}
\label{fig:jetregions}
\end{figure}

The \chandra\/ data used here are from the \source\ Very Large Project
(\source-VLP), comprising six deep ($\sim 100$~ks) exposures (OBSIDs
7797, 7798, 7799, 7800, 8489, 8490) of the front-illuminated CCD chips
of the Advanced CCD Imaging Spectrometer (ACIS) made between 2007
March 7 and May 30. Standard
methods\footnote{http://asc.harvard.edu/ciao/} have been followed for
data re-processing and grade selection, and small astrometric
corrections were made to register the six exposures to a common frame
using detected point sources \citep{hard07, jordan07}.  The 0.8-3~keV
merged image of the inner $\sim 4'$ of the X-ray jet is shown in
Figure~\ref{fig:jet}.

We are interested here in the jet after it has become radiatively
bright in both radio and X-ray \citep[more than $21''$ from the core,
and after knots AX1A, AX1C in the terminology of ][]{hard03b} until
its major decline relative to the radio (at $\sim 141''$).  Motivated
by \ntof, where the fast spine is approximately 60\% of the jet width,
we have defined regions shown in Figure~\ref{fig:jetregions}.  While
not wishing to suggest {\it a priori\/} that the \source\ jet does
have a slower sheath and faster spine, for shorthand we call these
regions the {\it spine}, {\it sheath} and {\it background}, working
outwards from the jet axis.  Compact sources excluded from the
background regions are also marked.  A division of the spine and
sheath into two regions with length down the jet is shown.  We refer
to these as the {\it inner} and {\it outer} regions, and the boundary
was chosen so that there is a comparable number of total counts in
each region.

The jet is positioned on a single ACIS I chip in all six exposures.
The boundary between the outer and inner regions lies between 1.4 and
4.0 arcmin off axis depending on OBSID, giving a point-spread-function
(PSF) half-power-diameter (HPD) at this location of better than about
1 arcsec at the $\sim 1.5$-keV energy of maximum count rate per keV.
The readout streak from the \source\ core runs close to the SE sheath
region in OBSID 7797, and so we exclude this observation in
spectral fitting to the SE sheath and inner spine.
We extract data and responses for identical sky regions in the
different OBSIDs, and we fit the same model simultaneously to the data
sets.  Spectra are binned to a minimum of 30 counts per bin so that
the $\chi^2$ statistic can be applied.

In this work we are searching for statistical trends in spectral index
in knotty and diffuse emission transverse to the jet axis. Knots are
identified as bright enhancements, but we believe them to be composed
of multiple distinct physical structures, each potentially with a
different emission spectrum.  The precise definition of knots is
therefore not crucial, and we have adopted the previously published
lists of \citet{kraft02} and \citet{kataoka}, which combined give 31
knots within our total spine and sheath regions.  The knots were
examined visually, and small centroid shifts (within uncertainties)
were made to give best alignment in the new data.  A circle of nominal
radius 2.4 arcsec was used to mask each knot.  Four additional weak
but compact enhancements seen in the new data were also masked as
knots with circles of radii 1.2 arcsec.  Because of our interest in
statistical trends, for knots that overlap regions we divide the
counts according to where they were detected.  When extracting spectra
of the knots the diffuse (residual) emission in the corresponding
region is used as background.  The outer sectors shown in
Figure~\ref{fig:jetregions} provide background for the diffuse
emission.

All our spectral models are single-component power laws with free
hydrogen column density, $N_{\rm H}$, over the energy range 0.4--10
keV.  Intrinsic $N_{\rm H}$ in \source\ over and above the Galactic
value of $8.4 \times 10^{20}$ cm$^{-2}$ \citep{dlock90} causes
reductions in the count rate in soft X-ray images, consistent with the
dust lanes that cross the jet roughly transverse to its axis.  Since
our regions extend hundreds of pc along the jet, we are combining data
with different intrinsic $N_{\rm H}$.  This is expected to give a bias
towards finding slightly flatter power-law slopes than the true slope
if the jet emission is characterized by a single spectrum over its
length. However, the gradients in $N_{\rm H}$ are principally down the
jet and not across the jet, and so provide similar weightings when we
compare sections transverse to the jet axis.  We can thus use the
fitted $N_{\rm H}$ and spectral index for our transverse sections of
the jet as proxies of systematic trends in the spectra.  Appreciable
cross-jet differences could then allow us to conclude that there are
real variations in the X-ray spectrum across the jet.
Self-consistency tests (see \S\ref{sec:results}) verify this
assertion.  We also find, through comparison with the spectra of
individual knots and sections of diffuse emission along the jet
\citep{hard07}, that any bias in spectral slope in our analysis is
small.

In the following, spectral index $\alpha$ (one less than X-ray photon
index) is defined in the sense that flux density is proportional to
$\nu^{-\alpha}$.

\begin{deluxetable*}{clccccc}
\tablewidth{0pt}
\tablecaption{\chandra\ fits to a single-component power law
\label{tab:specresults}}
\tablehead{
\colhead{Row} & \colhead{Component/} &
\colhead{region \% counts} &
\colhead{region \% area} &
\colhead{$\alpha$\tablenotemark{a}}  & 
\colhead{$N_{\rm H}$\tablenotemark{a}($10^{21}$ cm$^{-2}$)} & 
\colhead{$\chi^2_{\rm min}$/dof}
}
\startdata
1 & Spine & 100 & 100
& $0.83^{+0.05}_{-0.04}$ & $1.54^{+0.18}_{-0.17}$ & 961/895 \\
2 & Sheath& 100 & 100
& $0.97^{+0.08}_{-0.07}$  & $1.40\pm0.28$ & 527/512  \\
3 & NW sheath & 100 & 100
& $0.98^{+0.12}_{-0.11}$ & $1.47^{+0.46}_{-0.44}$ & 318/277 \\
4 & SE sheath & 100 & 100
& $0.99\pm0.11$ & $1.51^{+0.45}_{-0.42}$ & 295/282 \\
5 & Spine knots & 64 & 25
& $0.60\pm0.06$  & $1.77^{+0.29}_{-0.28}$ & 602/657 \\
6 & Sheath knots & 49 & 17
& $0.73^{+0.12}_{-0.11}$ & $1.12^{+0.53}_{-1.06}$ & 244/254 \\
7 & Spine diffuse & 36 & 75
& $1.24\pm0.10$  & $1.79^{+0.37}_{-0.34}$ & 471/407 \\
8 & Sheath diffuse & 51 & 83
& $1.20\pm0.14$  & $1.82^{+0.51}_{-0.50}$ & 338/308 \\
9 & Inner spine knots & 77 & 47
& $0.63\pm0.08$ & $2.34^{+0.44}_{-0.42}$ & 474/520 \\
10 & Inner sheath knots & 61 & 33
& $0.94^{+0.17}_{-0.18}$ & $1.64^{+0.80}_{-0.76}$ & 221/185 \\
11 & Outer spine knots & 34 & 18
& $0.98^{+0.17}_{-0.18}$ & $0.09^{+1.48}_{-0.06}$ & 209/207 \\
12 & Outer sheath knots & 27 & 13
& $0.80^{+0.30}_{-0.40}$ & $0.05^{+1.90}_{-0.05}$ & 60/71 \\
13 & Inner spine diffuse & 23 & 53
& $0.96^{+0.16}_{-0.17}$ & $2.85^{+0.83}_{-0.88}$ & 182/168 \\
14 & Inner sheath diffuse & 39 & 67
& $0.90^{+0.25}_{-0.22}$ & $2.73^{+1.27}_{-1.16}$ & 136/121 \\
15 & Outer spine diffuse & 66 & 82
& $1.35^{+0.12}_{-0.11}$ & $1.46^{+0.40}_{-0.36}$ & 318/336 \\
16 & Outer sheath diffuse & 73 & 87
& $1.33^{+0.19}_{-0.17}$ & $1.41^{+0.63}_{-0.59}$ & 191/191 \\
17 & ``Bright'' spine knots & 88 & 57
& $0.74^{+0.11}_{-0.10}$ & $1.36^{+0.51}_{-0.48}$ & 229/249 \\
18 & ``Bright'' sheath knots & 82 & 60
& $0.89^{+0.23}_{-0.19}$ & $1.12^{+0.96}_{-0.90}$ & 147/124
\enddata
\tablenotetext{a}{Errors are 90\% for 2 interesting parameters}
\end{deluxetable*}

\section{Results}
\label{sec:results}

\begin{figure}
\centering
\includegraphics[width=0.71\columnwidth,clip=true]{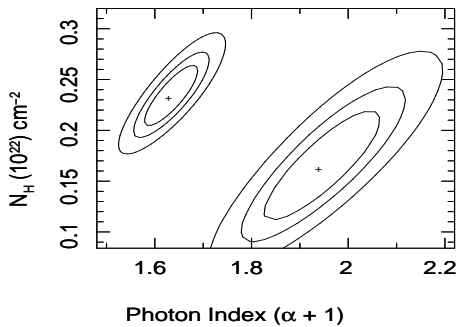}
\caption{Confidence contours ($1\sigma$, 90\% and 99\% for two
  interesting parameters) showing a significant difference between
the spine inner knots (left contours)
  and sheath inner knots (right contours): rows 9 \& 10 of
  Table~\ref{tab:specresults}.  The $y$ axis begins at Galactic
  $N_{\rm H}$.
}
\label{fig:contours}
\end{figure}

\begin{figure}
\centering
\includegraphics[width=0.71\columnwidth,clip=true]{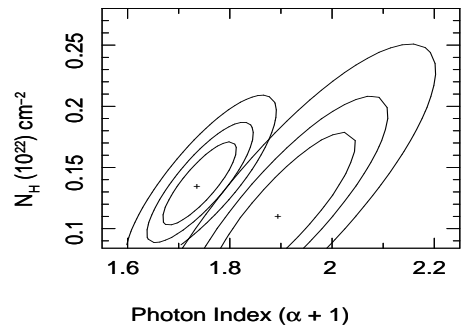}
\caption{Same as Fig.~\ref{fig:contours} for the
bright region spine knots (left contours)
  and sheath knots (right contours): rows 17 \& 18 of
  Table~\ref{tab:specresults}. 
}
\label{fig:contours2}
\end{figure}

Our spectral results are summarized in Table~\ref{tab:specresults},
and we refer to them here by row number.  The quoted uncertainties are
90\% for two interesting parameters, so where pairs of rows disagree
in both parameters there is at most a 1\% probability of agreement by
chance.  Uncertainties in $\alpha$ and $N_{\rm H}$ are correlated, and
so the $\chi^2$ contours must be examined to assess fully the spectral
disagreement. We show such plots only for the comparisons of most
interest (Figs.~\ref{fig:contours} \& \ref{fig:contours2}).

We see (Table~\ref{tab:specresults} rows 1 \& 2) that the sheath
spectrum is steeper than the spine.  Further investigation allows us
to find the origin of this result.  The NW and SE regions of the
sheath (rows 3 \& 4) give excellent agreement in $\alpha$ and $N_{\rm
H}$, indicating that it is unlikely that the result is from isolated
features, and confirming the absence of strong gradients in $N_{\rm
H}$ transverse to the jet.  Separating the knot and diffuse emission
in the spine and sheath (rows 5 to 8) shows that while the knots have
flatter spectra than the diffuse emission, the steeper spectrum in the
sheath is caused solely by differences in the knots.  In the
two-dimensional plane of $\alpha$ and $N_{\rm H}$, rows 5 and 6
disagree.

Having established that the knots in the sheath have a steeper
spectrum than in the spine, we checked the inner and outer regions to
see if there is any trend with distance from the core.  Comparing rows
9 and 10 with 11 and 12, we see the expected preference for a higher
$N_{\rm H}$ in the inner regions, but more interestingly we see that
it is only for the knots in the inner regions that the spine and
sheath spectra differ (Fig.~\ref{fig:contours}).  The diffuse
emission, in contrast, shows remarkable spectral agreement between the
sheath and spine in the inner region (rows 13 \& 14).  To check that
the contour separation in Figure~\ref{fig:contours} is causal rather
than statistical we have drawn random sets of the required size from
the modelled underlying spectral-index distribution deduced from all
knots in the inner region.  The average spectral index has an error on
the mean of $\pm 0.07$ for knots matching in number those in the inner
spine, and $\pm 0.11$ for the inner sheath.  The observed
spectral-index difference, $\Delta\alpha = 0.31$
(Fig.~\ref{fig:contours}), therefore significantly exceeds the
difference expected from random subsamples of the distribution.

Since the spine inner jet knots tend to lie in regions of higher
$N_{\rm H}$ than the sheath inner knots (Figs~\ref{fig:jetregions} and
\ref{fig:contours}), we have repeated the analysis only for the
``bright'' region, marked in blue on Figure~\ref{fig:jetregions}.
This region is beyond the main dust lanes that cross the jet
\citep[e.g., fig.~1 of][]{hard07}.  Results are consistent with sheath
knots having steeper spectra than spine knots
(Fig.~\ref{fig:contours2}, and rows 17 \& 18).  Simulations show that
the bias (\S 2) on the absolute value of mean spectral index from the
$N_{\rm H}$ distribution over the inner jet is $\sim 0.05$.

\begin{figure}
\centering
\includegraphics[width=1.0\columnwidth,clip=true]{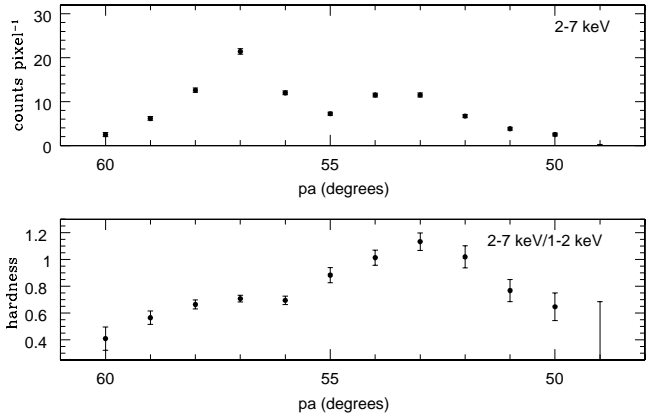}
\caption{Surface-brightness profiles with position angle
of the knot emission from the inner jet in a hard energy
band (upper panels) with the hardness ratio below.  The X-ray
emission is softer
towards the jet edges
($\chi^2 = 116$ for 11 dof for a fit to constant hardness).  There is
no trend for regions of higher integrated knot surface brightness to
show harder spectra.
}
\label{fig:inknotprofile}
\end{figure}

We investigated the possibility that the spectral flattening in the
inner spine might result from the inner knots being brighter, if
brightness is associated with particle-acceleration efficiency.
Figure~\ref{fig:inknotprofile} shows the knot surface brightness and
hardness across the inner jet.  The bin-width is 1 degree in position
angle, which corresponds roughly to the HPD at an average jet-axis
distance, and so the points are not highly correlated.  The hardness
ratio shows the decrease with increasing distance from the jet axis
that is anticipated from our spectral results, with a high degree of
confidence that the hardness is not constant.  The gradual decrease of
hardness with distance from the jet axis shows that our finding of a
spectral difference in the spine and sheath regions was not dictated
by the {\it a priori\/} definition of the region boundaries.

Perhaps surprisingly, a comparison of the panels of
Figure~\ref{fig:inknotprofile} shows that there is no trend for
regions of higher integrated knot surface brightness to be harder in
spectrum.  We have made plots similar to
Figure~\ref{fig:inknotprofile} for the inner and outer diffuse
emission, and the outer knot emission.  The surface-brightness
profiles of diffuse emission show some lumpiness, suggesting that some
of what we call diffuse emission is composed of
lower-surface-brightness knots, but all three hardness-ratio profiles
are consistent with being constant across the jet ($\chi^2$ values of
10.5, 7.9, and 14.8 for 11 degrees of freedom (dof), as compared with
$\chi^2$ =116 in Fig.~\ref{fig:inknotprofile}).

\section{Conclusions}
\label{sec:conclude}

If the knots in \source\ were predominantly a surface feature in a
shear layer between the spine and sheath, as we suggested might be the
case for \ntof, one would expect the knot X-ray spectra to be similar
across the jet.  This is not what is seen.  In the inner jet, between
$21''$ and $66''$ from the core, we have observed in a statistical
sense that the X-ray knot spectra steepen farther from the jet axis.
No such trend is seen in the more diffuse X-ray emission.  The results
suggest that knots are distributed throughout the jet volume, with a
trend towards steeper spectra in the outer regions.  The trend will
then be underestimated in our results because of projection effects.

Although the spectral trends disfavor knot confinement in a shear
layer, they imply that the electron energy distributions in knots
towards the center and edge of the jet are different. Electrons with
sufficient energy to emit synchrotron X-rays must be accelerated
locally to the knots, since their energy loss timescale is short
compared with any dynamical timescale associated with fluid motion.
Thus the conditions for particle acceleration differ across the
jet. The knots might arise from intruders from outside travelling at
speeds close to the stellar velocity dispersion, or internal flow
irregularities which are most likely to arise in the central part of
the flow and might be expected to move quickly towards the edge of the
jet under the influence of circulation in the shear layer.  In either
case a transverse velocity structure might cause the strength of
shocks near the knots to vary across the jet, and since the kinetic
energy density of the flow should be higher nearer the jet axis, we
might also expect velocity irregularities to generate stronger
turbulent cascades near the axis than near the jet edge, with
consequent efficient particle acceleration \citep{manolakou} and a
flatter on-axis X-ray spectrum.

As the jet expands, the lower on-axis flow velocity, as suggested by
the less collimated appearance of the jet further from the core, would
then imply less X-ray spectral difference between sheath and spine,
consistent with our results.  Indeed, a spectral steepening between
the inner and outer regions in diffuse X-ray emission may relate to
such a decline in flow speed along the jet (rows 13 to 16 of
Table~\ref{tab:specresults}).

It is interesting to ask whether it will be possible to detect knot
migration in \source.  Expansion of the jet along its axis, with
apparent motions of about $0.5 c$, has been measured in radio data
from the Very Large Array over 10 years \citep{hard03b}, but lateral
motion is more difficult to measure.  It has been detected in the
radio in M\,87 \citep{biretta95}.  We speculate that transverse
motions should be present in the knots and are potentially detectable
if knots are flow irregularities, although future observations may be
required.

These new observations of \source\ have permitted the first X-ray
spectral study of a radio jet in the transverse direction, separating
knotty and more diffuse emission.  We have found the striking result
that where the X-ray emission is brightest, at $\sim 0.4$ to 1.2 kpc
from the core, the spectrum in the knotty emission is steeper at
larger distances from the jet axis.  We speculate that this is the
result of lateral knot migration.
\smallskip


This work was partially supported by NASA grant GO7-8105X.
We thank the referee for constructive comments and MJH
thanks the Royal Society for support.
Facilities: \facility{CXO}

\end{document}